# SCIENTIFIC AND TECHNOLOGICAL DEVELOPMENT OF HADRONTHERAPY

SAVERIO BRACCINI[*]

*Albert Einstein Centre for Fundamental Physics,
Laboratory for High Energy Physics (LHEP), University of Bern,
Sidlerstrasse 5, CH-3012 Bern, Switzerland*

Hadrontherapy is a novel technique of cancer radiation therapy which employs beams of charged hadrons, protons and carbon ions in particular. Due to their physical and radiobiological properties, they allow one to obtain a more conformal treatment with respect to photons used in conventional radiation therapy, sparing better the healthy tissues located in proximity of the tumour and allowing a higher control of the disease. Hadrontherapy is the direct application of research in high energy physics, making use of specifically conceived particle accelerators and detectors. Protons can be considered today a very important tool in clinical practice due to the several hospital-based centres in operation and to the continuously increasing number of facilities proposed worldwide. Very promising results have been obtained with carbon ion beams, especially in the treatment of specific radio resistant tumours. To optimize the use of charged hadron beams in cancer therapy, a continuous technological challenge is leading to the conception and to the development of innovative methods and instruments. The present status of hadrontherapy is reviewed together with the future scientific and technological perspectives of this discipline.

## 1. Introduction

Hadrontherapy – often also denominated 'particle therapy' - is a collective word used to indicate the treatment of tumours through external irradiation by means of accelerated hadronic particles. Several kind of particles have been and are the subject of intensive clinical and radiobiological studies: neutrons, protons, pions, antiprotons, helium, lithium, boron, carbon and oxygen ions. Among all these possibilities, only two of them – protons and carbon ions – are nowadays used in clinical practice and represent the focus of an ongoing remarkable scientific and technological development. For this reason, only proton and carbon ion therapy are discussed in this paper.

Protons and carbon ions are more advantageous in cancer radiation therapy with respect to X-rays mainly because of three reasons. The release of energy along their path inside the patient's body is characterized by a large deposit localized in the last few millimetres at the end of their range, in the so called Bragg peak region, where they produce severe damage to the cells while sparing both traversed and deeper located healthy tissues. Moreover, they penetrate the patient with minimal diffusion and, using their electric charge, few millimetre FWHM 'pencil beams' of variable penetration depth can be precisely guided towards any part of the tumour. The third reason pertains to carbon ions - and light ions in general - and is based on radiation biology. Since, for the same range, carbon ions deposit about a factor 24 more energy in the Bragg peak region with respect to protons, the produced ionization column is so dense to be able to induce direct multiple strand brakes in the DNA, thus leading to non-repairable damage. This effect is quantified by an enhancement of the Radio Biological Effectiveness (RBE) and opens the way to the treatment of tumours, which are resistant to X-rays and protons at the doses prescribed by standard medical protocols.

---
[*] E-mail: Saverio.Braccini@cern.ch.

In order to treat deep-seated tumours, depths of the order of 25 cm in soft tissues have to be reached. This directly translates into the maximum energies of proton and carbon ion beams which must be 200 MeV and 4 500 MeV (i.e. 375 MeV/u), respectively. As far as the beam currents are concerned, the limit is set by the amount of dose to be delivered to the tissues, typically 2 Gy per litre per minute. This translates into beam currents on target of 1 nA and 0.1 nA for protons and carbon ions, respectively.

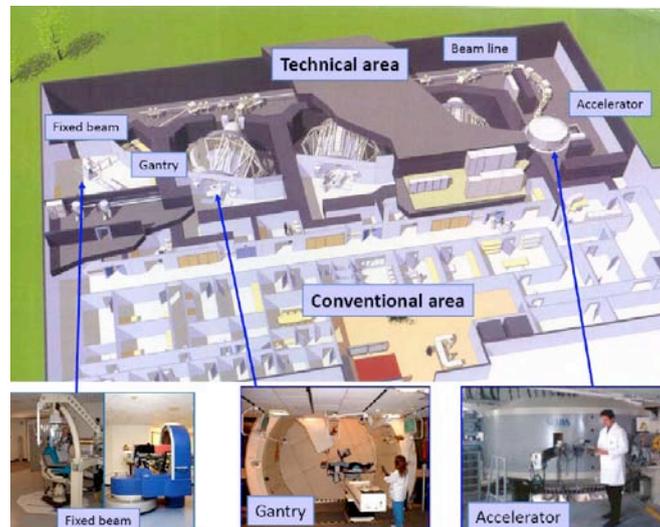

Figure 1. General layout of a proton therapy centre featuring one accelerator, three treatment rooms equipped with rotating gantries and one room equipped with a fixed horizontal beam. The example reported here is based on a system commercialized by the company IBA (Belgium).

The maximum energy and current determine the main characteristics of the accelerator: 4-5 metre diameter cyclotrons, both at room temperature and superconducting, and 6-8 metre diameter synchrotrons are today in use for proton therapy while for carbon ion therapy only 20-25 metre diameter synchrotrons are available.

As presented in Fig. 1, a modern hadrontherapy centre is based on a complex facility in which the accelerator is connected to several treatment rooms by means of beam transport lines. The treatment rooms are equipped with fix beams – usually horizontal but vertical and 45˚ are also in use – or rotating gantries, which, for proton therapy, are about 10 m high, 100 tons structures supporting a set of magnets, able to irradiate the patient from any direction, exactly like in conventional X-ray radiation therapy.

The first idea of using accelerated protons and ions in cancer radiation therapy dates back to 1946 when Bob Wilson wrote a very illuminating seminal paper[1] in which all the basic principles and potentialities of this discipline are stated. I personally find this work very remarkable and still incredibly actual, especially if one considers the fact that precise imaging techniques and enough powerful accelerators were almost a dream at that time.

## 2. Present status of proton and carbon ion therapy

The first proton therapy treatment took place in Berkeley[2] in 1954, followed by Uppsala in 1957. This pioneering work opened the way to the intensive activity performed at the Harvard cyclotron where physicists and radiation oncologists worked together for many decades on three clinical studies: neurosurgery for intracranial lesions (3 687 patients), eye tumours (2 979 patients) and head-neck tumours (2 449 patients). The results obtained by the Harvard group represented the basis for the successive clinical and technical developments of this discipline.

A fundamental milestone was accomplished in 1990 when the first patient was treated at the Loma Linda University Medical Center in California, the first hospital based proton therapy centre. This facility featured the first rotating gantries designed for routine treatment. It has to be remarked that, up to this moment, all the hadrontherapy facilities were based on existing particle accelerators designed for fundamental research, often sharing human resources and beam time with other activities. Moreover, some of these centres made use of low energy – about 70 MeV – cyclotrons, in which only the treatment of ocular pathologies was possible.

In the last twenty years, a progressive development of proton therapy took place. From being practiced only in specialized research nuclear and particle physics laboratories, proton therapy is becoming a widely recognized clinical modality in oncology. As reported in Fig. 2, many hospital based centres are nowadays active in the world.

| Centre | Country | Acc. | Max. Clinical Energy (MeV) | Beam Direction (a) | Start of treat. | Total treated patients | Date of total |
|---|---|---|---|---|---|---|---|
| ITEP, Moscow | Russia | S | 250 | H | 1969 | 4 024 | Dec-07 |
| St.Petersburg | Russia | SC | 1000 | H | 1975 | 1 327 | Dec-07 |
| PSI, Villigen (b) | Switzerland | C | 72 | H | 1984 | 5 076 | Dec-08 |
| Dubna (c) | Russia | SC | 200 | H | 1999 | 489 | Dec-08 |
| Uppsala | Sweden | C | 200 | H | 1989 | 929 | Dec-08 |
| Clatterbridge (b) | England | C | 62 | H | 1989 | 1 803 | Dec-08 |
| Loma Linda | USA | S | 250 | 3 G, H | 1990 | 13 500 | Dec-08 |
| Nice (b) | France | C | 65 | H | 1991 | 3 690 | Dec-08 |
| Orsay (d) | France | SC | 200 | H | 1991 | 4 497 | Dec-08 |
| iThemba Labs | South Africa | C | 200 | H | 1993 | 503 | Dec-08 |
| MPRI(2) | USA | C | 200 | H | 2004 | 632 | Dec-08 |
| UCSF (b) | USA | C | 60 | H | 1994 | 1 113 | Dec-08 |
| TRIUMF, Vancouver (b) | Canada | C | 72 | H | 1995 | 137 | Dec-08 |
| PSI, Villigen (e) | Switzerland | C | 250 | G | 1996 | 426 | Dec-08 |
| HZB (HMI), Berlin (b) | Germany | C | 72 | H | 1998 | 1 227 | Dec-08 |
| NCC, Kashiwa | Japan | C | 235 | 2 G, H | 1998 | 607 | Dec-08 |
| HIBMC, Hyogo | Japan | S | 230 | 2 G, H | 2001 | 2 033 | Dec-08 |
| PMRC(2), Tsukuba | Japan | S | 250 | 2 G, H | 2001 | 1 367 | Dec-08 |
| NPTC, MGH, Boston | USA | C | 235 | 2 G, H | 2001 | 3 515 | Oct-08 |
| INFN-LNS, Catania (b) | Italy | C | 60 | H | 2002 | 151 | Dec-07 |
| Shizuoka | Japan | S | 235 | 2 G, H | 2003 | 692 | Dec-08 |
| WERC, Tsuruga | Japan | S | 200 | H, V | 2002 | 56 | Dec-08 |
| WPTC, Zibo | China | C | 230 | 3 G, H | 2004 | 767 | Dec-08 |
| MD Anderson Cancer Center, Houston, TX (f) | USA | S | 250 | 3 G, H | 2006 | 1 000 | Dec-08 |
| FPTI, Jacksonville, FL | USA | C | 230 | 3 G, H | 2006 | 988 | Dec-08 |
| NCC, Ilsan | South Korea | C | 230 | 2 G, H | 2007 | 330 | Dec-08 |
| RPTC, Munich (g) | Germany | C | 250 | 4 G, H | 2009 | treatments started | Mar-09 |
| TOTAL | | | | | | 50 879 | |

(a) Horizontal (H), vertical (V), gantry (G).
(b) Ocular tumours only.
(c) Degraded beam.
(d) 3676 ocular tumours.
(e) Degraded beam for 1996 to 2006; dedicated 250 MeV proton beam from 2007. Scanning beam only.
(f) With spread and scanning beams (since 2008).
(g) Scanning beam only.

Figure 2. Hospital based proton therapy facilities[3] in operation at the end of 2008.

More that 50 000 patients have been treated by proton therapy worldwide, very often affected by pathologies located near critical structures, the so called organs at risk (OAR). It is important to remark the importance of proton therapy for paediatric patients, for whom the irradiation of critical organs may lead to severe permanent consequences on the quality of life and to possible induction of secondary cancers.

Many proton therapy centres are nowadays under construction or in a phase of advanced project and more that ten new hospital based facilities are expected to be operational in the next five years, mostly located in US, Europe and Japan. It is important to acknowledge the important contribution to this dvelopment given by the main commercial companies active in this field: Optivus (USA), IBA (Belgium), Varian/Accel (USA/Germany), Hitachi (Japan) and Mitsubishi (Japan).

Carbon ion therapy is still on its way to become a commonly accepted clinical option. Due to its specificity and to the complex radiobiological models essential for treatment planning, this irradiation modality is nowadays facing a phase of intense clinical and technical research.

Since the construction of the Heavy Ion Medical Accelerator in Chiba (HIMAC), which treated the first patient in 1994, Japan is the most advanced country in carbon ion therapy. By the end of 2009, about 5000 patients have been treated and many radio resistant tumours have been shown to be controllable.

In Europe, treatments started at the 'pilot project' for carbon ion therapy at GSI in Germany in 1997. Since then, about 400 patents have been treated and this project lead to the construction of the first European hospital based centre in Heidelberg (HIT). This new centre features a gigantic 600 tons, 25 metre long carbon ion gantry – the only one constructed so far - and treated the first patient in November 2009. In Italy, the Centro Nazionale di Adroterapia Oncologica (CNAO) is in a phase of very advanced construction and will be the second hospital based centre in Europe. Two more centres are under construction in Germany by the company Siemens and two projects, Etoile in France and MedAustron in Austria have been approved.

In a few years from now, several hospital based centres for carbon ion therapy will be operational and will produce the necessary knowledge needed to exploit at best the clinical potential of this highly ionizing radiation.

## 3. Future perspectives

Since a few years, hadrontherapy is facing a deep change which is bringing a relatively small scientific community into a much larger multi disciplinary contest in which physics, biology, engineering, medicine, law, management and finance came together and play all an important role. This is mainly due to the large size of this kind of projects which, being of the order of 100 million Euro, have a big impact not only on the scientific and technological side but also on the financial, logistic and management aspects. Needles to say that in a complex project, such as the construction of a hadrontherapy centre, many details have to be carefully evaluated and a solid multi-competence project team represents the key to face all the unavoidable problems and compromises to be assessed from the conception to the running of the facility.

In my opinion, two main forces – not always pointing towards the same direction - are driving the developments of this discipline. On one hand, to be competitive with the continuously increasing performances of conventional radiation therapy, innovative tools and techniques are needed to exploit at best the higher potential of hadron beams. On the other hand, to be able to offer this treatment modality to a larger number of patients, the cost, the size and the complexity of the equipment have to be reduced.

To face these challenges, many scientific and technological developments are ongoing and many new ideas are appearing at the horizon. A non-exhaustive summary of some selected topics is reported here.

### 3.1. *Dose delivery systems*

The dose delivery system represents a key issue in the full treatment process since allows the transformation of the beam coming out from the accelerator into a clinical three-dimensional dose distribution which has to comply with the medical prescription.

Up to present and in almost all the centres, the 'passive spreading' method - presented in Fig. 3 – is used. Although relatively simple, this method requires a large amount of mechanical work and does not allow exploiting at best the ballistic properties of charged hadron beams. In particular, extra doses are produced at the edges of the tumour.

In the last fifteen years, two innovative techniques have been developed in two research laboratories in Europe to 'paint' the tumour by means of a small 'pencil beam' driven by magnetic forces and without the use of any passive device, as presented in Fig. 4. The 'spot scanning'[4] and the 'raster scanning'[5] allow obtaining very conformal dose distributions and, by the superposition of non-uniform dose distributions coming from many directions, they allow very complex treatments by Intensity Modulated Particle Therapy (IMPT). Scanning techniques are nowadays becoming mature and commercial companies are offering this modality, often in parallel with the standard passive spreading.

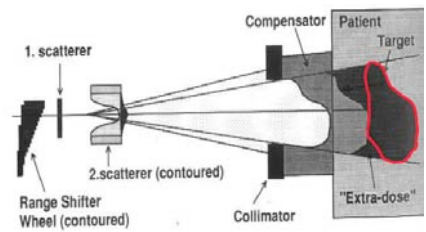

Figure 3. In 'passive spreading', the beam is widened and flattened by means of scatterers and adapted to the volume to be irradiated by means of personalized collimators and compensators. Different penetration depths are superimposed by means of a range shifter to obtain the Spread Out Bragg Peak (SOBP) covering the tumour.

To fully exploit the potentialities of scanning and IMPT, a new dedicated gantry is now under commissioning at PSI. The 'Gantry 2'[6] allows scanning with parallel beams in a two-dimensional surface and rotates only of 180˚ to allow an easier access to the patient with a more compact structure. This innovative device futures also a robotic couch for patient positioning and advanced imaging modalities. The possibility to use X-rays parallel to the proton beam and passing through a special aperture in the 90˚ magnet is foreseen to optimize the treatment of moving organs.

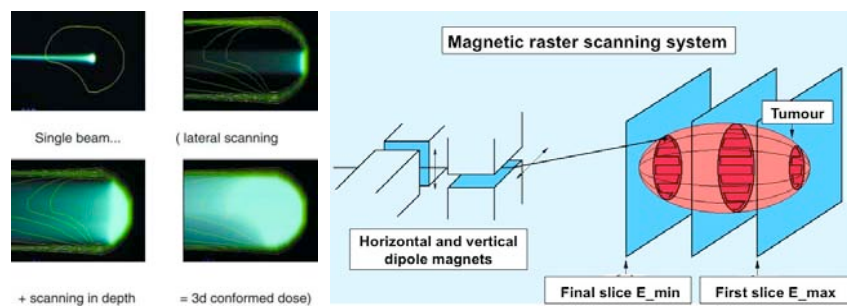

Figure 4. Developed at PSI, the 'spot scanning' technique consists on chopping the beam from a cyclotron to obtain discrete dose spots of given energy and direction which are used to form a uniform dose distribution (left). Developed at GSI, the 'raster scanning' consists on the use of a pencil beam to paint a slice of the tumour; the irradiation of successive slices, corresponding to different beam energies, gives a three-dimensional uniform dose distribution (right).

It has to be remarked that most of the tumours treated so far by particle therapy – like for example head and neck cancers - are not subject to motion due to respiration. The treatment of moving organs, like in the case of lung cancer, represents an important challenge and a very actual theme of research. In Japan, a synchronization 'gating' between the beam and the patient's respiration is already used and the irradiation takes place only when the tumour is in the appropriate position. More sophisticated techniques of active compensation are under study at GSI where the realisation of an on-line feedback system will allow the change of position and energy of the beam to correct organ motion effects.

Due to the much higher magnetic rigidity, a conventional magnet ion gantry is necessarily very big, like the one constructed at HIT. To make carbon ion gantries more compact, the use of superconductivity is under study together with more 'exotic' beam transport techniques such as the Fixed Field Alternating Gradient (FFAG). Non-isocentric gantries are also proposed. In this case the movement of the beam is limited – and therefore the number of magnets – which implies moving the patient to perform irradiations from any direction.

### 3.2. *Imaging and quality assurance*

In modern radiation therapy imaging is essential and computed tomography (CT) is the basis of treatment planning. In the near future, the role of MRI and PET functional imaging will surely be more important. In particular, innovative molecules[7] for PET imaging are nowadays in a phase of advanced study and will provide fundamental information for treatment planning on hypoxia and on the localization of tumours which are almost invisible to CT.

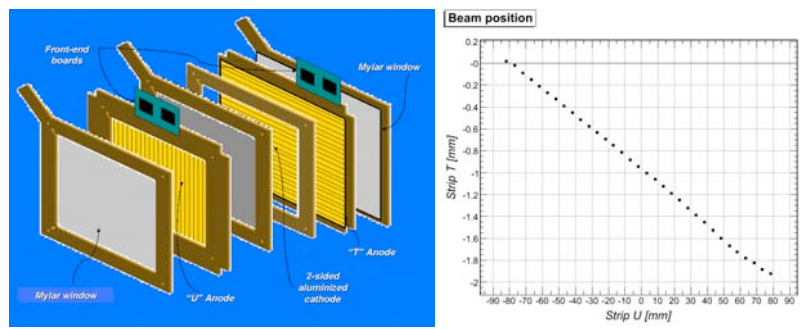

Figure 5. SAMBA (Strip Accurate Monitor for Beam Applications) is composed by two 2 mm strip ionization chambers (left). This detector is capable to monitor on-line the beam intensity and the transverse position with sub-millimetric precision during spot scanning (right).

On-line monitoring of hadron beams during treatment is of paramount importance for quality assurance. Many interesting developments in the field of particle detectors have been performed in the last years as the innovative strip ionization chamber[8] developed by the TERA Foundation and INFN-Torino for the Gantry 2 (Fig. 5).

In radiation therapy, the control of the dose distribution relies on off-line dosimetry and not on direct on-line measurements. At GSI, an innovative technique[9] – the in-beam PET – has been developed exploiting the fact that carbon ions form $^{11}$C nuclei by interacting with the patient's body (Fig. 6). Since $^{11}$C is a positron emitter, its distribution can be measured by means of a PET camera installed in the treatment room and compared with the treatment planning. Studies are underway also to exploit this technique for proton-therapy[10].

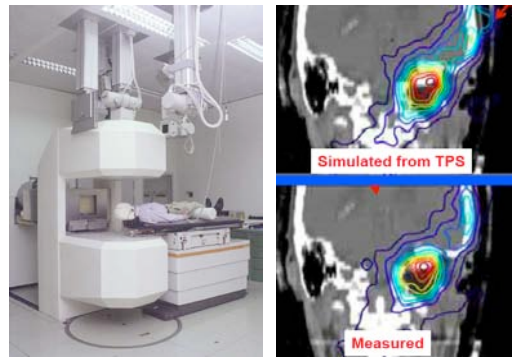

Figure 6. The treatment room at GSI equipped with the on-beam PET detector (left). Measured and simulated data demonstrate that the dose released by carbon ions is effectively conform with respect to the treatment planning.

The precise determination of the range is essential in proton therapy and, being based on CT data, correction factors have to be applied in the treatment planning to calculate the density of the traversed material. Proton radiography can be used to obtain direct information on the range for treatment planning optimization and to perform imaging with negligible dose to the patient. Proton radiography is an interesting theme of research and detectors based on silicon trackers and nuclear film emulsions are proposed[11].

### 3.3. *Particle accelerators*

Contrary with respect to conventional radiation therapy, where each treatment room is equipped with a linac, only one radiation source is available in a hadrontherapy facility. For this reason, the choice of the particle accelerator is of paramount importance and its up-time represents a very critical issue.

At present, four commercial companies offer proton accelerators - IBA and Varian/Accel offer cyclotrons, Hitachi and Mitsubishi synchrotrons - while for carbon ion therapy the only commercial solution is the synchrotron proposed by Siemens.

Cyclotrons are characterized by a beam of fixed energy which has to be degraded with passive absorbers in the first part of the beam line – in the so called Energy Selection System (ESS) – and, with respect to the human breathing cycle, present a continuous beam, advantageous for the treatment of moving organs.

On the other hand, the beam of a synchrotron presents a cycle – the so called 'spill' – which lasts about 2 seconds and in which the beam is present for about 0.5 seconds. This time structure is similar to the human respiratory cycle, posing problems for the treatment of moving organs. The energy can be varied from spill to spill and no passive absorbers are needed.

To reduce size, cost and complexity, single room facilities look to be a very promising solution, which may lead to a deep change of the entire discipline in the next future. The company Still River System is proposing a 10 Tesla superconducting synchrocyclotron to be directly installed in a 180˚ rotating gantry (Fig. 7 – left). The construction of the first prototype is in a very advanced phase and it surely represents a very interesting development to follow. With a longer research and development timescale, the company TomoTherapy, together with the Lawrence Livermore National Laboratories, is proposing a single room system based on an innovative accelerating technique, the Dielectric Wall Accelerator (DWA). Studies are underway to prove that beams suitable for proton therapy can be obtained by means of this new technology.

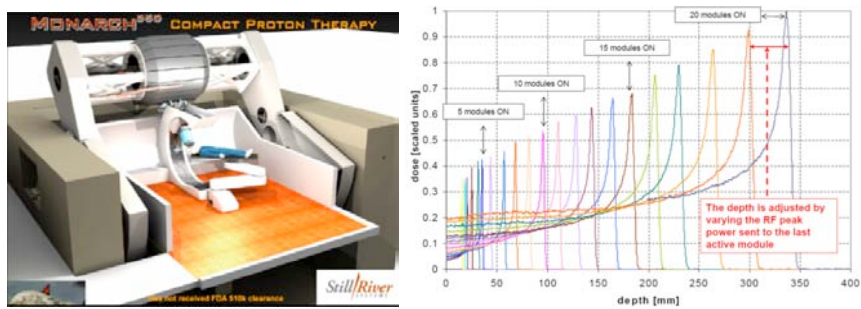

Figure 7. The single room proton therapy facility proposed by Still River Systems (left). With a linac the beam energy can be continuously changed by switching off the last modules and by varying the power in the last active module (right).

Radio frequency linacs[12] represent a very interesting solution since they naturally allow very fast changes of the beam energy in a time scale of 1 ms, consistent with the repetition rate of the accelerator. As presented in Fig. 7 (right), thanks to the rapid 'bunch-to-bunch' energy variation, linacs are ideal radiation sources for spot scanning.

Initiated by the TERA Foundation, a system based on a high current cyclotron followed by a linac post accelerator is proposed by the company ADAM. This solution allows proton therapy and radioisotope production with a single accelerator complex.

For carbon ion therapy, the company IBA is studying an innovative multi-particle superconducting cyclotron and the first prototype of this machine is under construction in Caen (France). A similar 300 MeV proton and 300 MeV/u superconducting cyclotron named SCENT is proposed by INFN-LNS and IBA and will allow full treatments in proton therapy and carbon ion therapy up to a maximum penetration of 16 cm in water. A linear post accelerator named CABOTO is proposed by the TERA Foundation with the scope of boosting carbon ions to the energy of 400 MeV/u to treat any kind of tumour.

Looking into the very far future, the acceleration of protons and ions by means of laser plasma devices represents a fascinating field of research[13]. It has to be remarked that, at present, only the maximum energy of 67.5 MeV for protons has been recently achieved. Moreover, protons are produced with a very wide energy spread peaked at low energies, making of the dose distribution system a very difficult challenge.

## 4. Conclusions

Since its beginnings, particle physics has always offered medicine and biology tools and techniques to study, detect and cure cancer. Hadrontherapy represents today one of the major contributions of particle physics to the medical filed and is now facing a very exciting phase at the forefront of science and technology.

## Acknowledgements

I would sincerely acknowledge Ugo Amaldi who first introduced and than guided me into this fascinating field of application of particle physics to medicine. I would like to thank Vesna Sossi for having given me the possibility to present hadrontherapy at the 11[th] ICATPP Conference in Como.